
\font\titolino=cmbx10
\font\tsnorm=cmr10
\font\tscors=cmti10
\font\tsnote=cmr9

\font\tsnotec=cmti9
\magnification=1200
\hsize=148truemm 
\hoffset=10truemm
\parindent 8truemm
\parskip 3 truemm plus 1truemm minus 1truemm
\nopagenumbers
\newcount\notenumber

\def\note{\advance\notenumber by 1 \footnote{$^{\the\notenumber}$}}
\def\beginref{\begingroup\bigskip
\leftline{\titolino References.}
\nobreak\smallskip\noindent}
\def\endref{\par\endgroup}
\def\ref#1#2{\noindent\item{\hbox to 25truept{[#1]\hfill}} #2.\smallskip}
\def\beginsection #1. #2.
{\bigskip
\leftline{\titolino #1. #2.}
\nobreak\medskip\noindent}
\def\beginappendix #1.
{\bigskip
\leftline{\titolino Appendix #1.}
\nobreak\medskip\noindent}

\def\bn{\bigskip\noindent}
\def\sc{\scriptstyle}
\def\scc{\scriptscriptstyle}
\def\bh{bla\-ck \-ho\-le}

\def\RN{Reis\-sner -- Nord\-str\"om}
\def\Sc{Sch\-warz\-sch\-ild}
\def\Schr{Sch\-r\"o\-din\-ger}

\def\ffi{\varphi}
\def\pa{p_a}
\def\pb{p_b}
\def\pf{P_{\sc\Phi}}
\def\pphi{p_\varphi}
\def\d{\partial}
\def\ra{\rightarrow}
\def\h{\hat}
\def\de{\Delta_{FP}}

\def\frac#1#2{{{#1}/over{#2}}}
\def\AP#1#2#3{{\tscors Ann.\ Phys.} {\bf #1}, #2 (#3)}
\def\AJP#1#2#3{{\tscors Am.\ J.\ Physics} {\bf #1}, #2 (#3)}
\def\PR#1#2#3{{\tscors Phys.\ Rev.} {\bf #1}, #2 (#3)}
\def\PRD#1#2#3{{\tscors Phys.\ Rev.} {\bf D#1}, #2 (#3)}

\def\NPB#1#2#3{{\tscors Nucl.\ Phys.} {\bf B#1}, #2 (#3)}

\def\JMP#1#2#3{{\tscors J.\ Math.\ Phys.} {\bf #1}, #2 (#3)}

\def\IJMPA#1#2#3{{\tscors Int.\ J.\ Mod.\ Phys.} {\bf A#1}, #2 (#3)}
\def\IJMPD#1#2#3{{\tscors Int.\ J.\ Mod.\ Phys.} {\bf D#1}, #2 (#3)}

\def\CQG#1#2#3{{\tscors Class.\ Quantum Grav.} {\bf #1}, #2 (#3)}
\def\GRG#1#2#3{{\tscors Gen.\ Rel. Grav.} {\bf #1}, #2 (#3)}

\null
\vskip 5truemm
\rightline { }
\rightline{February 29, 1996}
\vskip 12truemm
\centerline{\titolino QUANTIZATION OF A 2D MINISUPERSPACE}
\bigskip
\centerline{\titolino MODEL IN DILATON-EINSTEIN GRAVITY}
\vskip 12truemm
\centerline{\tsnote MARCO CAVAGLI\`A$^\star$}
\smallskip
\centerline{\tsnotec SISSA - International School for Advanced Studies,}
\smallskip
\centerline{\tsnotec Via Beirut 2-4, I-34013 Trieste, Italy,}
\smallskip
\centerline{\tsnotec and}
\smallskip
\centerline{\tsnotec INFN, Sezione di Torino, Italy.}
\bigskip
\centerline{\tsnote VITTORIO DE ALFARO$^\dagger$} 
\smallskip
\centerline{\tsnotec Dipartimento di Fisica Teorica dell'Universit\`a di
Torino,} 
\smallskip
\centerline{\tsnotec  Via Giuria 1, I-10125 Torino, Italy,}
\smallskip
\centerline{\tsnotec  and}
\smallskip
\centerline{\tsnotec INFN, Sezione di Torino, Italy.}
\vfill
\centerline{\tsnorm ABSTRACT}
\begingroup\tsnorm\noindent
We investigate a minisuperspace model of Einstein gravity plus dilaton
that describes a static spherically symmetric configuration or a
Kantowski -- Sachs like universe. We develop the canonical formalism and
identify canonical quantities that generate rigid symmetries of the
Hamiltonian. Quantization is performed by the Dirac and the reduced
methods. Both approaches lead to the same positive definite Hilbert
space. 
\vfill
\leftline{\tsnorm PACS: 04.40.Nr, 04.60.Ds, 04.50.+h.\hfill}
\smallskip
\hrule
\smallskip\noindent
\leftline{$^\star$ E-Mail: CAVAGLIA@SISSA.IT\hfill}
\leftline{$^\dagger$ E-Mail: VDA@TO.INFN.IT\hfill}
\endgroup
\vfill
\eject
\footline{\hfill\folio\hfill}
\pageno=1
\noindent
Scalar fields are a very important ingredient in the quantum treatment of
gravitation. Indeed, the low energy limit of the string models provides
dilaton and moduli fields, that seem to play an essential role in quantum
cosmology and black hole physics. So the consideration of systems formed
by the gravitational and scalar fields is not academic but is part of our
physical picture. Recently a great deal of attention has been dedicated
to the coupling of scalar to gravity, even in presence of gauge fields
[1]. 

The first part of this note investigates the canonical formulation of a
minisuperspace reduced model for gravity interacting with a massless
scalar field. This model describes two independent geodesically complete
spacetimes: one is a Kompaneets - Chernov - Kantowski - Sachs [2] (KS for
brevity) like universe, the other one is a static asymptotically flat
spacetime: a complete universe with a naked singularity, corresponding in
the limit of vanishing dilaton to the external of a \Sc\ \bh. While the
classical solutions are well known since long time [3], the quantum
theory has not been fully discussed. Then it is worthwile to investigate
the quantization for this model. This is the subject of the second part
of the paper. We will use an approach that we have proposed and applied
to the case of the spherically symmetric metric devoid of matter (\Sc) or
containing an electric field (\RN) [4]. 

Let us spend some words about the method. We introduce the canonical
formalism, integrate the gauge equations off the constraint shell, and
evidence the role of a set of gauge invariant canonical quantities. They
generate a group of rigid symmetries of the gauge equations and are
fundamental for the quantization of the theory that is performed by the
usual rules according to the Dirac method. The range of integration of
the variables in the non gauge fixed inner product is defined by their
classical range and the measure is determined by the requirement that it
be invariant under the rigid transformations. Then the gauge fixing \`a
la Faddeev -- Popov is implemented. This allows to define a positive
definite inner product in the gauge fixed Hilbert space and completes the
quantization of the system [5]. A different approach to quantization
consists in introducing a classical canonical identity (a second class
constraint) that allows to reduce the phase space to the gauge fixed,
physical subspace, and then quantize (unitary gauge). The canonical
identity expresses the coordinate parameter as a function of the
canonical variables -- coordinates and momenta -- and determines the form
of the Lagrange multiplier [6]. As in [4], the two methods lead to the
same final result and one obtains a gauge fixed Hilbert space with
positive definite norm. 

We start from the action for the four--dimensional low energy string
effective theory written in the Einstein frame (see for instance [7]) 
$$S={1\over 16\pi}\int d^4x\sqrt{-g}\left[R+2\Lambda-2\sigma
\d_\mu\varphi\d^\mu\varphi\right] \,,\eqno(1)$$
where $R$ is defined as in [8] and $\sigma=\pm 1$. We consider the
negative sign of sigma in order to take into account contributions from
moduli fields derived from the compactified manifold. 

The reduced Ansatz for the line element is
$$ds^2=-4a(\xi)d\eta^2+4n(\xi)d\xi^2+b^2(\xi)d\Omega^2\,,\eqno(2)$$
and correspondingly the ansatz for the scalar field is
$\varphi=\varphi(\xi)$. A priori we do not restrict the sign of  the
coefficients $a$, $b$, $n$ that constitute the metric and are our
Lagrangian variables. To start with we do not fix the gauge of the
coordinates, namely we do not set any connection between $\xi$ and the
Lagrangian variables $a$, $b$, $n$. This will be done later, when
quantizing, and we will see that a suitable gauge fixing in each one of
the two approaches to quantization (Dirac versus reduced space) allows to
define  the same positive definite Hilbert space. Since the volume
element of the metric (2) is $\sqrt{-g}=4b^2\sqrt{an}$, the signature is
always lorentzian. 

Using the Ansatz for the metric and scalar field, the action (1) becomes
(actually, this is the action density in $\eta$, integrated over
$d\Omega$) 
$$S=\int d\xi~ 2l\left[{\dot a b\dot b\over l^2}+{a\dot
b^2\over l^2}- \sigma{ab^2\dot\varphi^2\over l^2}+{1\over
4}\left(1+\Lambda b^2\right)\right]\,,\eqno(3)$$
where $l=4\sqrt{an}$ is a non--dynamical variable (Lagrangian multiplier)
and dots represent differentiation with respect to $\xi$. By a Legendre
transformation we obtain the Hamiltonian 
$${\cal H}\equiv l{\cal
H}_l={l\over 2b^2}\left[\pa\left(b\pb-a\pa\right)-\sigma\pphi^2/4a-b^2
\left(1+\Lambda b^2\right)\right]\,.\eqno(4)$$
The constraint deriving from the presence of the Lagrange multiplier is
thus ${\cal H}_l=0$ that is a direct consequence of the invariance of the
theory under reparametrizations of the parameter $\xi$. 

Let us now develop the canonical formalism and obtain the general
solutions. We put for the moment $\Lambda=0$. We will see later that the
absence of horizon and the completeness of the spaces hold also when
$\Lambda$ is different from zero. We define gauge invariant quantities
along the lines of our treatment of the \Sc\ \bh\ [4]. These quantities
form an interesting algebra that will be at the basis of the next
discussion of the quantization. 

The gauge transformations generated by $\cal H$ (denoted as ${\cal H}_h$)
can be integrated explicitly. From the gauge equations it follows that
$\pphi$ and $N=b\pb-2a\pa$ are gauge invariant quantities. Then we have
to discuss separately different cases, depending on the value of $\pphi$
and $N$. Let us define the quantity $H={\cal H}_l+1/2$ which reduces to
1/2 on the gauge shell. In the following we define $\tau=\int l(\xi)d\xi$
and consider without loss of generality $\tau>0$ and $H>0$. We now give
the general solution of the gauge equations according to the different
cases. 
\bn
{\bf Case 1}: $\gamma^{-2}=1+\sigma(\pphi^2/N^2)$ (this is certainly true
for $\sigma=1$) and $a\geq0$ (this corresponds to a static and isotropic
space -- time): 
$$\pa={N\over a}\left[{H\over N}\tau-{1\over 
2}\left(1+{1\over\gamma}\right)\right]\,,~~~~~~
\pb={N\over b}\left[{2H\over 
N}\tau-{1\over\gamma}\right]\,,$$
$$b={\tau\over 2I}\left[1-{N\over\gamma 
H\tau}\right]^{(1-\gamma)/2}\,,~~~~~~~~
a=2HI^2\left[1-{N\over\gamma H\tau}\right]^\gamma\,,\eqno(5)$$
$$\varphi=\Phi-\sigma{\pphi\gamma\over 2N}\ln{
\left[1-{N\over\gamma H\tau}\right]}\,,$$
where we consider for simplicity $\gamma>0$ since the solution is
invariant for $\gamma\ra -\gamma$ apart from a redefinition of $\tau$
($\tau\ra\tau-N/\gamma H$). $I$ and $\Phi$ are two gauge invariant
quantities defined as 
$$I=\sqrt{a\over 2H}\left[{\gamma b\pb+N\over\gamma 
b\pb-N}\right]^{\gamma/2}\,,~~~~~~
\Phi=\varphi+\sigma{\pphi\gamma\over 2N}
\ln{\left[{\gamma b\pb-N\over\gamma b\pb+N}\right]}\,.\eqno(6)$$
This solution corresponds to the well known solutions given in the
literature both for positive and negative $\sigma$ [1,3]. To see it, let
us fix the coordinate by going on the gauge shell, i.e.\ $H=1/2$, and
choosing $l=1$. Then $\tau=\xi$ and the coordinates $\xi$, $\eta$ can be
identified respectively with a radial and a timelike variable. The
solution becomes ($\xi\ra r$, $\eta\ra t$) 
$$\eqalign{&ds^2~=~-\left[1-{{2N}\over{\gamma r}}\right]^{\gamma}\,
dt^2 \,+
\left[1-{{2N}\over{\gamma r}}\right]^{-\gamma}\,dr^2\,+ r^2
\left[1-{{2N}\over{\gamma r}}\right]^{1-\gamma}\,d\Omega^2\,,\cr
&\ffi=\Phi-\sigma{\pphi\gamma\over 2N}
\ln{\left[1-{{2N}\over{\gamma r}}\right]}\,,\cr}\eqno(7)$$
where we have chosen $I=1/2$ so that $t$ is the proper time for
$r\rightarrow \infty$. Since $\gamma>0$, by inspection of (7) we can
easily see that for $r\to \infty$ the line element (7) is asymptotically
flat. The spacetime is defined for $r>r_s$, where $r_s=2N/\gamma$ for
$N>0$ and $r_s=0$ for $N<0$. In both cases, if $\gamma \not= 1$ $r=r_s$
is a naked curvature singularity (see for instance [9]). When $\gamma=1$,
i.e.\ $\varphi=$const., for $N>0$ (7) corresponds to the usual \Sc\
spacetime with mass $N$. In this case $r=r_s$ is a coordinate
singularity, so the spacetime can be continued for $r<r_s=2N$ down to
$r=0$, where there is a curvature singularity. This picture changes
completely when $\gamma\not=1$ because in this case at $r=r_s$ the area
of the two-sphere is vanishing or singular. Hence, the spacetime
described by (7) when $\gamma\not=1$ cannot be continued below $r_s$. 
\bn
{\bf Case 2}: $\gamma^{-2}=1+\sigma(\pphi^2/N^2)$ and $a\leq0$ (this
corresponds to a complete KS like universe): 
$$\pa={N\over a}\left[{H\over N}\tau-{1\over
2}\left(1+{1\over\gamma}\right)\right]\,,~~~~~~
\pb={N\over
b}\left[{2H\over N}\tau-{1\over\gamma}\right]\,,$$
$$b={\tau\over 2I}\left[{N\over\gamma
H\tau}-1\right]^{(1-\gamma)/2}\,,~~~~~~~~
a=-2HI^2\left[{N\over\gamma H\tau}-1\right]^\gamma\,,\eqno(8)$$
$$\varphi=\Phi-\sigma{\pphi\gamma\over 2N}\ln{ \left[{N\over\gamma
H\tau}-1\right]}\,,$$
where now
$$I=\sqrt{-a\over 2H}\left[{N+\gamma b\pb\over N-\gamma
b\pb}\right]^{\gamma/2}\,,~~~~~~
\Phi=\varphi+\sigma{\pphi\gamma\over 2N} \ln{\left[{N-\gamma b\pb\over
N+\gamma b\pb}\right]}\,.\eqno(9)$$
Since $a<0$ this is a complete KS like space -- time, $\xi$ is a timelike
coordinate and so the metric is time dependent. With the same choice of
the Lagrange multiplier and of $I$ and $H$ as in the previous case, and
setting $\xi\ra t$ and $\eta\ra \chi$ ($0\leq \chi < 2\pi$) in (8), the
solution takes the form 
$$\eqalign{&ds^2~=~-\left[{{2N}\over{\gamma t}}-1\right]^{-\gamma}dt^2\,
+\,\left[{{2N}\over{\gamma t}}-1\right]^{\gamma}\, d\chi^2 \,+\, t^2
\left[{{2N}\over{\gamma t}}-1\right]^{1-\gamma}\,d\Omega^2\,,\cr
&\ffi=\Phi-\sigma{\pphi\gamma\over 2N} \ln{\left[{{2N}\over{\gamma
t}}-1\right]}\,.\cr}\eqno(10)$$
Consider for simplicity $t>0$ and $N>0$. As in the case 1, when
$\gamma=1$ the line element coincides with the \Sc\ metric. When
$\gamma\not=1$, we have instead a curvature singularity at $t=t_s\equiv
2N/\gamma$. Hence, for $\gamma\not=1$ (10) represents a (complete)
anisotropic KS universe that begins in a curvature singularity at $t=0$
and ends at $t=t_s$ in a curvature singularity after a finite lapse of
time. Conversely, when the scalar field is absent, the metric (10)
reduces to standard \Sc\ solution and coincides (with a suitable
redefinition of coordinates) with the solution (7) for $\gamma=1$. In
particular, the solution (10) reduces, for vanishing scalar field, to the
internal \Sc\ region, and solution (7) to the external \Sc\ region. Since
the singularity in $t=t_s$ ($r=r_s$) is now a coordinate singularity,
both metrics can be continued across the horizon and so they coincide. 
\bn
{\bf Case 3}: $\gamma^{-2}=-[1+\sigma(\pphi^2/N^2)]$ (this case implies
$\sigma<0$ and does not allow having a pure \Sc\ \bh\ solution when the
dilaton is absent) and $a\geq0$: 
$$\pa={N\over a}\left({H\over N}\tau-{1\over 2}
\right)\,,~~~~~~
\pb={2H\tau\over b}\,,$$
$$b={|N|\over 4HI\gamma}\left[1+{4H^2\gamma^2\over N^2}\tau^2
\right]^{1/2}\exp\left[-\gamma~\hbox{\rm arctg}\;\left({2H\gamma\over
N}\tau\right)\right]\,,\eqno(11)$$
$$a=2HI^2\exp\left[2\gamma~\hbox{\rm
arctg}\;\left({2H\gamma\over N} \tau\right)\right]\,,~~~~~~
\varphi=\Phi-\sigma{\pphi\gamma\over N}\hbox{\rm arctg}\;
\left[{2H\gamma\over N}\tau\right]\,,$$
where
$$I=\sqrt{a\over 2H}\exp\left[-\gamma~\hbox{\rm
arctg}\;\left({b\pb\gamma\over N}\right)\right]\,,~~~~~~
\Phi=\varphi+\sigma{\pphi\gamma\over N} ~\hbox{\rm
arctg}\;\left({b\pb\gamma\over N}\right)\,.\eqno(12)$$
As in the previous cases, let us choose the Lagrange multiplier $l=1$. On
the gauge shell the solution becomes ($\xi\ra r$, $\eta\ra t$) 
$$\eqalign{ds^2=&-\exp\left\{2\gamma\left[{\rm
arctg} \left({2H\gamma\over N}r\right)-{\pi\over 2}\right]\right\}dt^2+\cr
&+\exp\left\{-2\gamma\left[{\rm arctg}
\left({2H\gamma\over N}r\right)-{\pi\over 2}\right]\right\}dr^2+\cr
&+{N^2\over\gamma^2}\left[1+{\gamma^2\over N^2}r^2
\right]\exp\left\{-2\gamma\left[{\rm arctg}\left({2H\gamma\over
N}r\right)-{\pi\over 2}\right]\right\}d\Omega^2\,,\cr\cr
\ffi=&\Phi-\sigma{\pphi\gamma\over N}\hbox{\rm arctg} \left[{\gamma\over
N}r\right]\,,\cr}\eqno(13)$$
where we have chosen $I=e^{-\gamma\pi/2}/2$.  As for the case 1 the
spacetime described by (13) is static and asymptotically flat in the
radial coordinate $r$. Further, it is never singular. Indeed, for $r\ra
0$ the line element becomes 
$$ds^2=e^{-\pi\gamma}\left\{-dt^2+e^{2\pi\gamma}\left[dr^2+
\left(N^2/\gamma^2\right)d\Omega^2\right]\right\}\,,\eqno(14)$$
and the scalar field assumes its minimum value. The area of the
two-sphere at $r=0$ has a finite value different from zero. This means
that the spacetime has a throat at $r=0$. Note that the existence of the
wormhole is made possible by the negative sign of $\sigma$. Indeed, in
this case the scalar field has negative energy density. The spacetime
(13) describes then a static traversable wormhole (see for instance
[10]). 

The quantities $\{N$, $I$, $\pphi$, $\Phi\}$ play a fundamental role. Let
us write their Poisson algebra: 
$${\vbox{
\offinterlineskip\halign{\strut\qquad#\qquad\hfil&
\qquad#\qquad\hfil&
\qquad#\qquad\hfil\cr
$[N,H]_P=0\,,$&$[I,H]_P=0\,,$&$[N,I]_P=I\,,$\cr
\noalign{\medskip}
$[\Phi,H]_P=0\,,$&$[\pphi,H]_P=0\,,$&$[\Phi,\pphi]_P=1\,,$\cr
\noalign{\medskip}
$[\Phi,N]_P=0\,,$&$[I,\Phi]_P=0\,,$&$[\pphi,I]_P=0\,,$\cr
\noalign{\medskip}
$[N,\pphi]_P=0\,.$&&\cr}}}\eqno(15)$$
Thus $N$ and $\ln I$, $\Phi$ and $\pphi$ are canonically conjugate
variables. Let us now introduce the new quantity $Y=(b\pb-a\pa)/H$, which
has the following Poisson parentheses: 
\medskip\noindent
\line{\hfill$[Y,I]=[Y,N]=[Y,\pphi]=[Y,\Phi]=0\,,$\hfill
$[Y,H]=1\,.$\hfill(16)}
\medskip\noindent
Thus a complete set of canonically conjugate variables are $\{N$,
$P_N\equiv\ln I$, $\Phi$, $\pf\equiv\pphi$, $Y$, $P_Y\equiv H\}$. Note
that all variables, except $Y$, are gauge invariant and generate rigid
transformations that leave invariant the Hamiltonian. This suggests
obviously to use $Y$ in order to fix the gauge (see later). Invariance
under rigid transformations will be used to determine the form of the
quantum measure.  Performing the canonical transformation to the new
variables the action becomes: 
$$S=\int d\xi\{\dot N P_N+\dot\Phi \pf+\dot Y
P_Y -l(P_Y-1/2)\}\,.\eqno(17)$$

In order to implement the Dirac procedure, the first main problem is the
choice of the variables to be used for the wave functions and of the
measure. The requirement of invariance of the measure under the rigid
transformations selects for instance the measure $d[\mu]=dp_N~dy~d\phi$.
(The eigenvalues of $P_N$, $Y$, and $\Phi$ have been indicated by lower
case letters). Given the measure, we have the representation of the
conjugate variables as differential operators: 
$${\vbox{
\offinterlineskip\halign{\strut\qquad#\qquad\hfil&
\qquad#\qquad\hfil&
\qquad#\qquad\hfil\cr
$\hat \pf\ra -i\d_{\sc\phi}\,,$&$\hat P_Y\ra -i\d_y\,,$
&$\hat N\ra i\d_{\sc p_{\scc  N}}\,,$\cr
$\hat\Phi\ra\phi\,,$&$\hat  Y\ra y\,,$
&$\hat P_N\ra P_N\,.$\cr}}}\eqno(18)$$
The WDW equation becomes
$$\left(-i\d_y-1/2\right)\Psi(y,p_N,\phi)=0\,.\eqno(19)$$
The solutions of (19) that are eigenfunctions of $\hat N$ and $\hat
\pf$ with eigenvalues $\nu$ and $\omega$ are
$$\Psi_{\nu,\omega}=C(\nu,\omega)\exp[-i\nu p_N+i\omega\phi+iy/2]\,.\eqno(20)$$

Now in order to progress we have to introduce the gauge fixing via the FP
method [5]. We will prove that there is a class of viable gauges for
which there are no Gribov copies and the FP determinant $\de$ is
invariant under gauge transformations. Indeed, let us suppose that the
gauge be enforced by $G(p_N,y,\phi)=0$, and let $G$ have the form 
$$G(p_N,y,\phi)=F(p_N,y,\phi)~\prod_i ~\bigl(y-g_i(p_N,\phi) \bigr)\,,
\eqno(21)$$
where $F\bigl(p_N,g_i(p_N,\phi),\phi\bigr) \not=0$ and
$g_i(p_N,\phi)\not=g_j(p_N,\phi)$ for any $p_N$ and $\phi$. Then we have
$$\de^{-1}~=~\int dh~\delta\bigl(G(h)\bigr)~=~\sum_i
f^{-1}_i(p_N,\phi)\,,\eqno(22)$$
where
$$f_i(p_N,\phi)= F\bigl(p_N, g_i(p_N,\phi),\phi\bigr)
 \prod_{j\not= i} \bigl(g_i(p_N,\phi)-g_j(p_N,\phi)\bigr)\,.\eqno(23)$$
Note that since $p_N$ and $\phi$ are gauge invariant, so is $\de$. The
gauge fixed invariant measure is then 
$$\int d[\mu]~ \delta\bigl(G(p_N,y,\phi)\bigr)~\de~=~\int
dp_N\, dy\,d\phi~
\delta\bigl(G(p_N,y,\phi)\bigr)~\de\,.\eqno(24)$$
In our case the most convenient gauge is $G(p_N,y,\phi)=y -
\xi=(bp_b-ap_a)/H-\xi=0$, where $\xi$ is a number. This gauge fixing
implies obviously $\de=1$ and determines uniquely the gauge. Now we may
discuss the form of the wave functions in this gauge. Denoting by lower
case greek letters the wave functions in the gauge fixed representation
and choosing $C(\nu)=(2\pi)^{-1}$, the gauge fixed eigenfunctions of $\h
N$ and $\h P_\Phi$ are 
$$\psi_{\nu,\omega}=
{1\over 2\pi}\exp[-i\nu p_N+i\omega\phi+i\xi/2]\eqno(25)$$
that are orthonormal. The eigenfunctions (25) coincide with those found
for the \Sc\ \bh\ in [4] apart from the plane wave in $\phi$. 

Let us now quantize the system by the alternative method of reducing
first the phase space by a canonical identity [6]. Again the  gauge
fixing condition is $Y=\xi$. This determines the Lagrange multiplier as
$l=1$ since from the definition of $Y$ and the classical general solution
of the gauge equations it follows $Y=\tau+{\rm constant}$. Using the
constraint $H=1/2$ and the gauge fixing condition, the effective
Hamiltonian on the gauge shell becomes $H_{\rm eff}=-P_Y=-1/2$. Thus the
\Schr\ equation is 
$$\left(i{\d~\over\d\xi}\;+\;{{1}\over{2}}\right)\;\psi~=~0\,.\eqno(26)$$
Diagonalizing $\h N$ and $\h P_\Phi$ we obtain the wave functions (25)
which form a definite positive Hilbert space. This proves the equivalence
of the Dirac and reduced quantization methods in the representation used.

Let us conclude by spending some words about the presence of cosmological
constant. We can deduce the properties of the solution without solving
the equations. The method we present here can also be applied, at least
in principle, to more complicate cases as for instance when a simple
potential term for the dilaton is present. 

Let us consider the action (3). If $\Lambda>0$, we can redefine the
Lagrange  multiplier as $l'(\xi)=l(\xi)f(b)$, where $f(b)=1+\Lambda b^2$.
The Lagrangian becomes 
$$L={f(b)\over l'}\left(2\dot a b\dot b+2a\dot b^2-2
\sigma ab^2\dot\varphi^2\right)+{l'\over 2}\,.\eqno(27)$$
Obviously the equations of motion from the Lagrangian (27) can be
interpreted as the geodesic equations for the minisuperspace manifold
$\tilde M$ with metric 
$$d\tilde s^2=f(b)[b~dadb+a~db^2-\sigma ab^2~d\varphi^2]\,.\eqno(28)$$
The solutions of the Einstein equations for the metric (2) with
cosmological constant are the geodesics of the minisuperspace. Thus we
can deduce the properties of the Einstein solutions from the study of the
geodesics of $\tilde M$. We are interested in establishing which points
in $\tilde M$ are singular. Indeed, no geodesic can cross a singular
point of $\tilde M$ and so this will be an end point of the solutions of
the Einstein equations for the metric (2). Note that the Einstein
solutions can be singular also for regular points of $\tilde M$, since
the metric of the minisuperspace is written in a given coordinate system
that can be pathological at some point. Thus this method does not give in
general all the curvature singularities of the space -- time (2);
however, if $\tilde M$ is singular, necessarily the space -- time (2)
must be singular. 

For the line element (28) the Kretschmann scalar is
$R_{\mu\nu\rho\sigma}R^{\mu\nu\rho\sigma}= 4(1+8\Lambda^2
b^4/f^2)/a^2b^4f^2$, so it is easy to see that there are no geodesics
such that the Kretschmann scalar is finite for $a=0$. This proves that
$a=0$ is a end point for all geodesics and that the space -- times for
$a>0$ and $a<0$ are separately complete. 

We are very grateful to Fernando de Felice and Alexandre T. Filippov for
interesting discussions and suggestions. 
\beginref

\def\AVA{M.\ Cavagli\`a, V.\ de Alfaro and A.T.\ Filippov,
\IJMPD{4}{661}{1995}}

\def\BEK{J.D.\ Bekenstein, \AP{82}{535}{1974}}

\def\BL{O.\ Beckmann and O.\ Lechtenfeld, \CQG{12}{1473}{1995}}

\def\BUC{H.\ Buchdal, \PR{115}{1325}{1959}}

\def\CDF{M.\ Cavagli\`a, V.\ de Alfaro, and A.T.\ Filippov, 
\IJMPA{10}{611}{1995}}

\def\CON{H.D.\ Conradi, \CQG{12}{2423}{1995}}

\def\DHS{D.\ Garfinkle, G.T.\ Horowitz, and A.\ Strominger, 
\PRD{43}{3140}{1991}} 

\def\GAVA{M.\ Cavagli\`a, V.\ de Alfaro, and A.T.\ Filippov, 
{\it Quantization of the \Sc\ Black Hole}, preprint DFTT 50/95, August
1995, gr-qc/95\-08\-062}

\def\GM {G.W.\ Gibbons and K.\ Maeda, \NPB{298}{741}{1988}}

\def\HT{For a review of the quantization of gauge systems, see M.\
Henneaux and C.\ Teitelboim, {\it Quantization of Gauge Systems}
(Princeton, New Jersey, 1992)}

\def\JRW {A.I.\ Janis, D.C.\ Robinson, and J.\ Winicour, 
\PR{186}{1729}{1969}}

\def\KAN{R.\ Kantowski and R.K.\ Sachs, \JMP{7}{443}{1966}}

\def\KC{A.S.\ Kompaneets and A.S.\ Chernov, {\it Zh.\ Eksp.\ Teor.\ Fiz.\
(J.\ Exptl.\ Theoret.\ Phys.\ (USSR))} {\bf 47}, 1939 (1964)} 

\def\LL{L.D.\ Landau and E.M.\ Lifshitz, {\it The Classical Theory of
Fields} (Pergamon Press, Oxford, 1975)}

\def\MAI{G.\ Lavreshlashvili and D.\ Maison, \NPB{410}{407}{1993}}

\def\MT{M.S.\ Morris and K.S.\ Thorne, \AJP{56}{395}{1988}}

\def\SIL{P.K.\ Silaev  and S.G.\ Turyshev, preprint hep-th/9510118} 

\def\TUR{S.G.\ Turyshev, \GRG{27}{981}{1995}}

\def\VJJ{K.S.\ Virbhadra, S.\ Jhingan and P.S.\ Joshi, preprint 
gr-qc/9512030}

\def\WYM{T.\ Wyman, \PRD{24}{839}{1981}}

\ref{1}{\WYM; \BL; \MAI; \TUR; \SIL; \CON}
\ref{2}{\KC; \KAN}
\ref{3}{\BUC; \JRW; \BEK}
\ref{4}{\AVA; \GAVA}
\ref{5}{\HT}
\ref{6}{\CDF}
\ref{7}{\DHS; \GM}
\ref{8}{\LL}
\ref{9}{\VJJ}
\ref{10}{\MT}

\endref

\bye